# Experimental demonstration of a free space cylindrical cloak without superluminal propagation

Su Xu[1,2,3], Xiangxiang Cheng[2,3], Sheng Xi[2,3], Runren Zhang[2,3], Herbert O. Moser[4], Zhi Shen[2], Yang Xu[2], Zhengliang Huang[5], Xianmin Zhang[2,3], Faxin Yu[5], Baile Zhang[6,7], and Hongsheng Chen*[1,2,3]

[1]*State Key Laboratory of Modern Optical Instrumentation, Zhejiang University, Hangzhou 310027, China*

[2]*Department of Information Science & Electronic Engineering, Zhejiang University, Hangzhou 310027, China*

[3]*The Electromagnetics Academy at Zhejiang University, Zhejiang University, Hangzhou 310027, China*

[4]*Network of Excellent Retired Scientists and Institute of Microstructure Technology, Karlsruhe Institute of Technology (KIT), Postfach 3640, D-76021 Karlsruhe, Germany*

[5]*School of Aeronautics and Astronautics, Zhejiang University, Hangzhou 310027, China*

[6]*Division of Physics and Applied Physics, School of Physical and Mathematical Sciences, Nanyang Technological University, Singapore 637371, Singapore*

[7]*Centre for Disruptive Photonic Technologies, Nanyang Technological University, Singapore 637371, Singapore*

We experimentally demonstrated an alternative approach of invisibility cloaking that can combine technical advantages of all current major cloaking strategies in a unified manner and thus can solve bottlenecks of individual strategies. A broadband cylindrical invisibility cloak in free space is designed based on scattering cancellation (the approach of previous plasmonic cloaking), and implemented with anisotropic metamaterials (a fundamental property of singular-transformation cloaks). Particularly, non-superluminal propagation of electromagnetic waves, a superior advantage of non-Euclidian-transformation cloaks constructed with complex branch cuts, is inherited in this design, and thus is the reason of its relatively broad bandwidth. This demonstration provides the possibility for future practical implementation of cloaking devices at large scales in free space.

*Authors to whom correspondence should be addressed; electronic mail: hansomchen@zju.edu.cn*

The current technology of invisibility cloaking is still facing serious bottlenecks after its tremendous development in the past few years. There are three major cloaking strategies so far, each of which has been developed almost independently without overlapping with others. The first strategy is the plasmonic cloaking based on scattering cancellation [1-3], where a plasmonic shell with negative permittivity can cancel the scattering from a dielectric dipole object. However, this strategy only applies to a specific dielectric object, and is fundamentally limited to subwavelength scales. The second strategy is based on a singular coordinate transformation [4-7] where a "hole" that can hide arbitrary objects is created in the electromagnetic space. The reason why this strategy does not violate the uniqueness theorem of inverse problems is because of the anisotropy of the designed cloak [5]. Although being able to cloak an arbitrary object, the cloak designed by this strategy is fundamentally narrowband because the speed of electromagnetic waves inside the cloak will exceed the speed of light in vacuum [8-10]. Although the recent carpet cloaking [11-18] can mitigate the bandwidth limitation by virtue of the support of a ground plane, cloaking a free-standing isolated object using this transformation method still suffers greatly from the superluminal limitation. The third strategy, utilizing transformation on a complex plane with properly designed branch cuts and Riemann sheets [19-20], can potentially get rid of superluminal velocity of electromagnetic waves by virtue of non-Euclidian transformation, but the required permittivity needs to continuously vary in a very large range [21] which is difficult to achieve in practice.

Although these strategies differ sharply in specific means and were developed almost independently without overlapping, they share the same target—to reduce scattering from the object to be hidden. Therefore, from a practical point of view with reverse thinking, a target-oriented approach might provide the possibility to integrate technical advantages of these strategies, and push further the current cloaking technology as a whole towards real applications. In this Letter, we experimentally demonstrated this possibility by designing and measuring a new type of cylindrical invisibility cloak in free space. First of all, the low scattering performance of the cloak is guaranteed by using the scattering cancellation method originated from plasmonic cloaking [1]. Instead of just cancelling the dipole moment as in plasmonic cloaking, our design method adopts a Mie scattering analytical model combined with an optimization algorithm to minimize the total scattering from the cloak [22]. Secondly, anisotropic metamaterials are used as building blocks to design and implement the cloak. Anisotropy is the key reason why the uniqueness theorem of inverse problems does not hold for singular-transformation cloaks [5]. The first singular-transformation cloak was fabricated by concentric multilayer anisotropic metamaterials [5]. Thirdly, our design intendedly imposes the condition of non-superluminal propagation of electromagnetic waves. Both permeability and permittivity are almost non-dispersive, and the effective refractive index has no component smaller than unit. The non-superluminal propagation, which was theoretically achievable in non-Euclidian transformation, is the mechanism underlying the relatively broad bandwidth of cloaking phenomena demonstrated in this work. By solving bottlenecks with integrated technical advantages, this alternative approach of invisibility cloaking may provide the potential to eventually implement a practical large-scale free-standing cloak in free space with broad bandwidth.

We start with the multilayer cylindrical cloak model (Fig. 1), which was adopted to implement the first singular-transformation cloak [5], and then integrate elements of other cloaking strategies based on it. For the sake of simplicity, we only consider a transverse electric (TE) polarized electromagnetic wave incident onto the cloak. The layered media are rotationally anisotropic in cylindrical coordinates. The constitutive parameters to be considered here are the relative permittivity in the z direction, $\varepsilon_z$, the relative permeability in the azimuthal direction, $\mu_\phi$ and the relative permeability in the radial direction, $\mu_\rho$. We use an optimization method that we proposed [22] to minimize the total radar cross section (RCS) of scattering. More details can be found in Supplementary Material. We impose the following constraints in the optimization for constitutive parameters: $1 < \varepsilon_z < 48$, $0 < \mu_\rho < 1$, $0.95 \leq \mu_\phi \leq 1$, and $\mu_\rho \varepsilon_z \geq 1.0$. The constraint $0 < \mu_\rho < 1$ indicates that we will use magnetic metamaterials, which will be explained in detail later. The constraint $\mu_\rho \varepsilon_z \geq 1.0$ is the condition of non-superluminal propagation inside the cloak. A small range of $0.95 \leq \mu_\phi \leq 1$ is introduced for testing the robustness of design, while in realistic implementation $\mu_\phi$ will always be unit. The frequency for optimization is set at 2.01 GHz. The object to be hidden is a cylinder of perfect electric conductor (PEC).

To compare with previous plasmonic cloaking which worked at subwavelength scales, we first design a one-layer cloak to hide a subwavelength object. The inner diameter of the cloak is $d = 29.8$ mm ($0.2\lambda$ at 2.01 GHz) and the outer diameter of the cloak is $D = 44.8$ mm （$0.3\lambda$ at 2.01 GHz）. The optimized constitutive parameters are: $\varepsilon_z = 24$, $\mu_\phi = 1$, and $\mu_\rho = 0.2$. Regarding the issue of experimental implementation, one of the reasons why previous cloaks that could cloak a PEC cylinder in free space [5] were strongly dispersive is because they rely heavily on split-ring-resonator (SRR) metamaterials, which are notoriously lossy and dispersive, to create magnetic responses. Here, for the first time, we use closed rings (CRs) [23, 24] as a non-resonant metamaterial to construct an almost dispersionless cloak. In each unit cell (as the inset in Fig. 2a), two metallic closed rings are printed on both sides of the substrate. It has been theoretically shown [23, 24] that the relative permeability of the closed rings is $\mu_\rho = 1 - F$, where $F = \frac{(a-w)(b-w)}{L_\phi L_z}$ is the fractional area of the periodic unit cell in $\phi z$ plane occupied by the interior of the closed rings. Therefore, we can design the dimensions of closed rings ($a$ = 6.9 mm and $b$ = 3.1 mm in the inset of Fig. 2a) to achieve required permeability. The required permittivity can be achieved by changing the substrate and adjusting the ratio between $a$ and $b$. In this design, we use FR4 (permittivity equals 4) as the substrate, with dimensions $L_\phi$ = 3.3 mm, $L_z$ = 7.8 mm, and $L_\rho$ = 0.5 mm. The width of the CR is $w$ = 0.2 mm and the thickness of the CR is $t$ = 0.034 mm. A retrieval exercise [25] shows that the effective permittivity and permeability of this CR metamaterial are $\varepsilon_z$ = 23.7, $\mu_\phi$ = 1, and $\mu_\rho$ = 0.22 from 1.6 to 2.4 GHz (Fig. S1b and c in Supplementary Material), very close to the required parameters. Note that this metamaterial composed of CRs is almost nondispersive over a wide frequency range, despite its permeability component significantly below unit. As explained in [23], this seemingly counterintuitive phenomenon does not violate causality. Moreover, the effective refractive index of this CR metamaterial has no component smaller than unit, i.e. $\mu_\phi \varepsilon_z \geq 1.0$ and $\mu_\rho \varepsilon_z \geq 1.0$ are satisfied. This property provides the possibility to achieve broadband cloaking in free space without superluminal propagation. Finally, the imaginary parts of constitutive parameters are nearly zero. This property is extremely desirable to construct a practical cloak with low loss. All of these superior advantages are difficult to achieve with traditional metamaterials.

Ten metamaterial unit cells of CRs in the $\rho$ direction and thirty unit cells of CRs in the $\phi$ direction are assembled to construct the one-layer cloak (Fig. 2a). Inside the cloak is the PEC cylinder. The scattered electric energy distributions in the near-field for the bare PEC cylinder (Fig.

2b) and for the cloaked PEC cylinder (Fig. 2c) are calculated. The scattered electric energy is reduced significantly around all the directions compared to the case of bare PEC cylinder. The experimental setup for measuring the RCS in all directions in the far-field from the PEC cylinder with and without the cloak is illustrated in Fig. 1. The differential RCS can be defined as $\sigma = 2\pi\rho \frac{|E_s|^2}{|E_i|^2}$, where $\rho$ is the distance from the center of the cloak to the receiver antenna [26], whose value is set to be 1 m in the experiment. $E_i$ and $E_s$ are the incident field at the object and the scattered field at distance $\rho$, respectively. The distance $R_i$ between the transmitter antenna and the cylinder is also set to be 1 m. The PEC cylinder with height $H = 152.6$ mm is used in the experiment. The measuring angle $\phi$ is varied from 0 degree (forward direction) to 160 degree. The measured results at 2.12 GHz that agree well with simulated results are shown in Fig. 2d.

Next, we define the ratio between the total RCS from the cloaked PEC cylinder and that from the bare PEC cylinder to evaluate the cloaking performance. Note that the angular range from 160 degree to 180 degree is not included because of the technical difficulty in measurement, which will slightly increase the error. Fig. 2e shows this RCS ratio at different frequencies calculated for the homogeneous ideal cloak and that measured for the constructed cloak. We can see that over a wide frequency band from 1.5 GHz to 2.3 GHz, the ratio is smaller than unity, which indicates that the cloak can reduce the scattering from the PEC cylinder over a broad frequency band. Even if a stricter criterion of 0.5 for the RCS ratio is imposed, there is still a bandwidth of 20% (from 1.79 GHz to 2.19 GHz). The maximum reduction of RCS up to about 90% for an arbitrary dipole object is already competitive with previous plasmonic cloaking that was applicable only to a specific dielectric object [2]. We have included in Supplementary Material another successful characterization of this one-layer cloak in a rectangular waveguide. Its three-dimensional performance in free space is also included in the Supplementary Material. Although the cloak essentially is a 2D cylindrical cloak, it still demonstrates valid cloaking effects for incident angles in the range of $\pm 15$ degrees.

After demonstrating a subwavelength cloak, we proceed to consider cloaking an object beyond subwavelength scales. When such a large cloaking area (*i.e.* beyond one-wavelength-large) is concerned, we need to use a multilayer cloak with more freedom to suppress more scattering modes [21], which will correspondingly take much longer computation time. For the sake of easier demonstration and faster computation, we adopted a six-layer invisibility cloak to cloak a PEC cylinder with diameter of $\lambda$. The optimized relative constitutive parameters in the order from the outermost layer to the innermost layer are $\mu_\rho = \{0.18, 0.36, 0.15, 0.15, 0.15, 0.15\}$, $\mu_\phi = \{1.00, 0.95, 1.00, 0.95, 1.00, 0.99\}$ and $\varepsilon_z = \{11.97, 38.85, 10.00, 36.30, 10.00, 19.15\}$. The thickness of each layer is $0.065\lambda$. All the parameters in the six layers are anisotropic, positive and finite. The refractive index components, i.e. $\sqrt{\mu_\phi \varepsilon_z}$ and $\sqrt{\mu_\rho \varepsilon_z}$, are greater than one. Therefore, the speed of electromagnetic waves traveling in the cloak is always slower than the speed of light in free space. This makes possible a large-size broadband invisibility cloak in free space. Note that the original cloak design with non-superluminal propagation was first proposed based on non-Euclidean transformation on a complex plane with proper branch cuts [21]. Our cloak is an alternative approach with similar effects of non-superluminal propagation.

With the constraints in optimization where $\mu_\phi$ in each layer is close to unity and $\mu_\rho$ is smaller than unity, this cloak can be realized with CRs in the manner similar to the previous subwavelength cloak. The CRs are fabricated on three different substrates: 0.5 mm-thick FR4, 0.25 mm-thick F4B and 0.025 mm-thick Polyimide (PI). More details are included in Supplementary Material. The retrieved parameters are very close to the required values. The photo of the implemented invisibility cloak is presented in Fig. 3a where the inset shows fabricated CR unit cells from Layer 1 to Layer 6. The height of the cylindrical cloak is $h = 230$ mm. The inner diameter of the cloak is $d_i = 149.25$ mm ($1\lambda$ at 2.01 GHz). The outer diameter of the cloak is $d_o = 266.15$ mm ($1.7832\lambda$ at 2.01 GHz). The calculated scattered energy distributions in the near field region of bare PEC cylinder and cloaked PEC cylinder at 2.01 GHz are compared in Fig. 3b and 3c. The measured far-field scattering from the bare PEC cylinder and that from the cloaked PEC cylinder in all directions are plotted in Fig. 3d, which agrees well with simulated values. The frequency for measurement is at 2.02 GHz. From both near-field and far-field results, we see the scattered field from the bare PEC cylinder is significantly suppressed by the cloak, especially the forward scattering (the reason of shadow behind a large object) that is reduced by more than 10 dB. We obtain the RCS ratio at different frequencies in Fig.3e. The ratio shows that our cloak can reduce RCS over the frequency range from 1.89 to 2.08 GHz, a span of 0.19 GHz or a 9.5% bandwidth. The maximum reduction in the total RCS up to 75.5% in experiment (82.7% in simulation), to the best of our knowledge, is the highest record in cloaking a one-wavelength-large PEC in free space. Due to the unique superior advantage of non-superluminal propagation, it is possible to extend this design to eventually cloak a large-scale object in free space with broad bandwidth.

Finally, it is interesting to discuss the reason why our alternative approach of cloaking based on an optimization procedure can work. Previous cloaking strategies can be categorized as "forward problems," where the properties of the cloak (such as the dispersion and the range of refractive index) can be worked out only after the design is done. That is why different strategies will lead to different cloaks with different properties. Our proposed approach, on the other hand, is one of the "inverse problems," where the properties of the cloak are predetermined with a proper range of optimization domain. The specific means and categories of design strategies (e.g. plasmonic cloaking, singular-transformation cloaking and non-Euclidian cloaking) are secondary from an integrated perspective. Apparently, the problem of this kind of inverse problems is that the existence of a proper solution is not always guaranteed. However, thanks to extensive studies on previous cloaking strategies which have accumulated plentiful knowledge, we know that some desired properties are achievable, such as broad bandwidth, non-superluminal propagation, and large-scale cloaking. These significant properties, although hard to achieve from individual previous strategies, may be possible for a target-oriented approach that can combine technical advantages of different strategies. After all, all current cloaking designs can be achieved in principle by a sufficiently powerful optimization program that can optimize all constitutive tensors independently. The inverse exploration from the desired properties by integrating previous strategies may provide a new prospect for the current cloaking technology.

In conclusion, we experimentally demonstrated an alternative approach of invisibility cloaking that can combine technical advantages of all current major cloaking strategies in a unified manner. A subwavelength cylindrical cloak and then a one-wavelength-large cylindrical cloak in free space are designed and measured. The design is based on scattering cancellation originated from previous plasmonic cloaking. Anisotropic metamaterials, a fundamental property of previous singular-transformation cloaking, are used to construct the cloaks. The superior advantages include non-superluminal propagation of electromagnetic waves in the cloak, previously only achievable with non-Euclidian transformation. These advantages may provide the possibility for future practical implementation of free-standing cloaking devices at large scales in free space.

**Acknowledgements**

This work was sponsored by the National Natural Science Foundation of China under Grants Nos. 61275183, 61274123, 60990320, and 60990322, the Foundation for the Author of National Excellent Doctoral Dissertation of PR China under Grant No. 200950, the ZJNSF under Grant No. R12F040001, and Singapore Ministry of Education under Grant No. MOE2011-T3-1-005.

**References**

1. A. Alu, and N. Engheta, Phys. Rev. E **72**, 016623 (2005).

2. A. Alu, and N. Engheta, Phys. Rev. Lett. **100**, 113901 (2008).

3. B. Edwards, A. Alu, M. Silveirinha, and N. Engheta, Phys. Rev. Lett. **103**, 153901 (2009).

4. J. B. Pendry, D. Schurig, and D. R. Smith, Science **312**, 1780 (2006).

5. D. Schurig *et al.*, Science. **314**, 997 (2006).

6. W. Cai, U. K. Chettiar, A. V. Kildishev, and V. M. Shalaev, Nat. Photonics **1,** 224–226 (2007).

7. H. Chen, B. -I. Wu, B. Zhang, and J. A. Kong, Phys. Rev. Lett. **99**, 063903 (2007).

8. Z. Ruan, M. Yan, C. W. Neff, and M. Qiu, Phys. Rev. Lett. **99**, 113903 (2007).

9. N. Kundtz, D. Gaultney, and D. R. Smith, New J. Phys. **12**, 043039 (2010).

10. H. Hashemi, B. Zhang, J. D. Joannopoulos, and S. G. Johnson, Phys. Rev. Lett. **104**, 253903 (2010).

11. J. S. Li, and J. B. Pendry, Phys. Rev. Lett. **101**, 203901 (2008).

12. R. Liu *et al.,* Science **323**, 366 (2009).

13. J. Valentine, J. Li, T. Zentgraf, G. Bartal, and X. Zhang, Nat. Mater. **8**, 568 (2009).

14. L. H. Gabrielli, J. Cardenas, C. B. Poitras, and M. Lipson, Nat. Photonics **3**, 461 (2009).

15. T. Ergin, N. Stenger, P. Brenner, J. B. Pendry, and M. Wegener, Science **328**, 337 (2010).

16. H. F. Ma, and T. J. Cui, Nat. Commun. **1**, 21 (2010).

17. B. Zhang, Y. Luo, X. G. Liu, and G. Barbastathis, Phys. Rev. Lett. **106**, 033901 (2011).

18. X. Z. Chen *et al.*, Nat. Commun. **2**, 176 (2011).

19. U. Leonhardt, Science **312**, 1777 (2006).

20. U. Leonhardt, and Tomas. Tyc, Science **323**, 110 (2009).

21. J. Perczel, T. Tyc, and U. Leonhardt, New J. Phys. **13,** 083007 (2011).

22. S. Xi, H. Chen, B. Zhang, B. –I. Wu, and J. A. Kong, Phys. Rev. B. **79**, 155122 (2009).

23. H. Chen, L. Huang, X. Cheng, and H. Wang, Prog. in Electromagn. Res. PIER **115**, 317 (2011).

24. J. B. Pendry, A. Holden, D. Robbins, and W. Stewart, IEEE Trans. Microwave Theory Tech. **47**, 11 (1999).

25. H. Chen *et al.*, Opt. Express **14**, 12944 (2006).

26. C. A. Balanis, Advanced Engineering Electromagnetics. John Wiley & Sons, New York, 1989.

**Figures:**

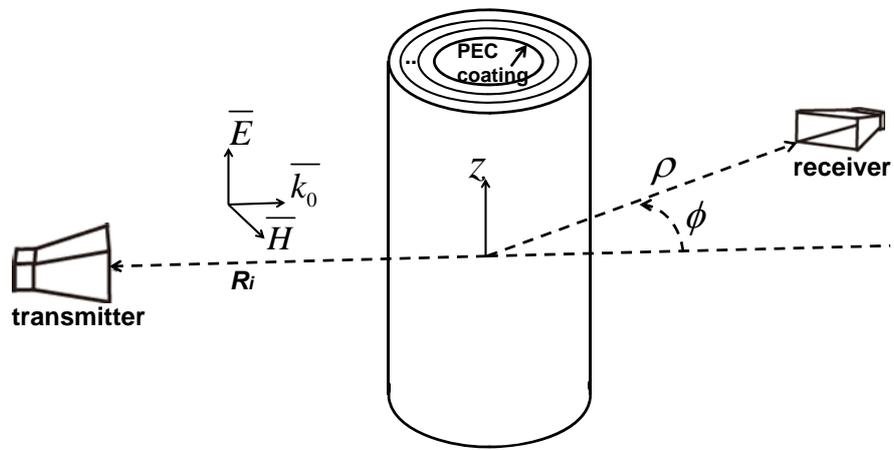

Fig. 1 Configuration of a multilayer cylindrical cloak and its experimental setup

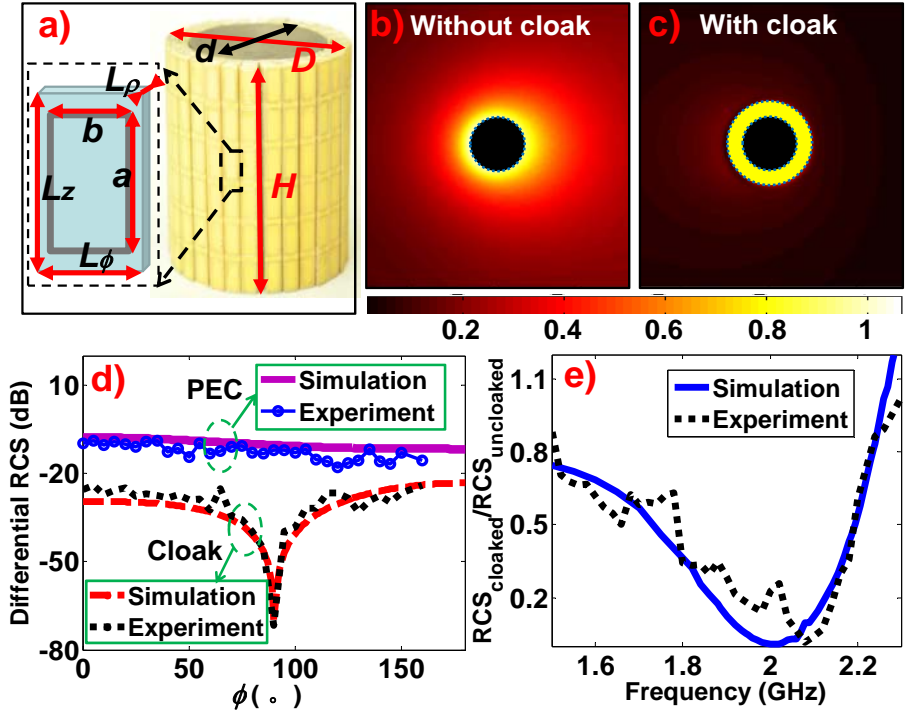

Fig. 2 (a) The subwavelength cylindrical cloak with periodic unit cells of closed rings printed on both sides of the substrate. The height, inner diameter and out diameter of the cloak are $H = 55$ mm, $d = 29.8$ mm and $D = 44.8$ mm, respectively. Parameters of the closed ring are: $a = 6.9$ mm and $b = 3.1$ mm. Dimensions of the substrate FR4 are $L_\phi = 3.3$ mm, $L_z = 7.8$ mm, and $L_\rho = 0.5$ mm. (b-c) The scattered electric energy distribution surrounding (b) the PEC core as indicated by the black circular area, and (c) the PEC with the cloak as indicated by the yellow annular area. (d) The experimental differential RCS via different azimuth angles $\phi$ at 2.12 GHz. (e) The reduction of RCS by the cloak in simulation (solid line) and in experiment (dotted line).

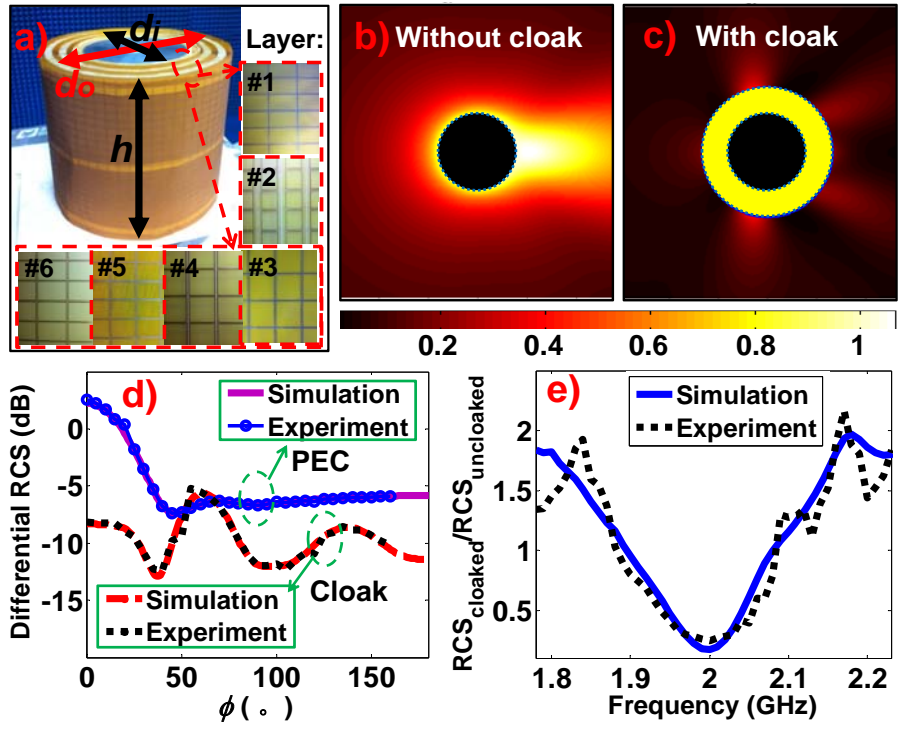

Fig. 3 (a) The six-layer cylindrical cloak with height $h$ =230 mm, inner diameter $d_i$ =149.25 mm (1$\lambda$ at 2.01 GHz), and outer diameter $d_o$ =266.15 mm (1.7832$\lambda$ at 2.01 GHz). The inset shows the fabricated CR cells from Layer 1 to Layer 6. (b-c) The calculated scattered electric energy distribution around (b) the bare PEC core as indicated by the black circular area, and (c) the PEC with the cloak as indicated by the yellow annular area. (d) The experimental RCS pattern via different azimuth angles $\phi$ at 2.02 GHz. (e) The reduction of total RCS by the cloak at different frequencies in simulation (solid line) and in experiment (dotted line).